\documentclass[aps,prb,floatfix,showpacs,superscriptaddress]{revtex4}
\usepackage{graphicx}
\usepackage{amsmath}
\begin{document}

\title{Switching the sign of photon induced exchange interactions in semiconductor
microcavities with finite quality factors}
\author{G. Chiappe}
\affiliation{Departamento de F\'{\i}sica Aplicada,
Instituto Universitario de Materiales and
Unidad Asociada del CSIC,
Universidad de Alicante, San Vicente del Raspeig, Alicante 03690, Spain.}
\affiliation{Departamento de  F\'{\i}sica J.J. Giambiagi, Facultad de
Ciencias Exactas, Universidad de Buenos Aires, Ciudad Universitaria,
1428 Buenos Aires, Argentina.}
\author{E. Louis}
\affiliation{Departamento de F\'{\i}sica Aplicada,
Instituto Universitario de Materiales and
Unidad Asociada del CSIC,
Universidad de Alicante, San Vicente del Raspeig, Alicante 03690, Spain.}
\author{E.V. Anda}
\affiliation{Departamento de  F\'{\i}sica, Pontificia Universidade
Cat\'olica do Rio de Janeiro (PUC-Rio), 22452-970, Caixa Postal:
38071 Rio de Janeiro, Brazil.}
\date{\today}

\begin{abstract}
We investigate coupling of localized spins  in a semiconductor
quantum dot embedded in a  microcavity with a finite quality factor.
The lowest cavity mode and the quantum
dot exciton are coupled  forming a polariton, whereas  excitons  interact
with localized spins via  exchange. The finite quality of the cavity $Q$ is
incorporated in the
model Hamiltonian by adding an imaginary part to the photon frequency. The
Hamiltonian,   which treats photons, spins and excitons  quantum
mechanically,  is solved exactly. Results for  a single  polariton clearly
demonstrate
the existence of a resonance, sharper as the temperature decreases, that shows up
as an abrupt change between ferromagnetic and antiferromagnetic
indirect {\it anisotropic} exchange  interaction between localized spins.
The origin of this spin-switching finite-quality-factor effect is discussed in detail
remarking on its dependence on model parameters, i.e., light-matter coupling,
exchange interaction between impurities, detuning and quality factor.
For parameters corresponding to the case
of a (Cd,Mn)Te  quantum dot, the resonance shows up for $Q \approx 70$ and detuning
around 10 meV. In addition, we show that, for such a quantum dot, and
the best cavities actually available (quality factors better than 200)  the exchange interaction
is scarcely affected.
\end{abstract}
\maketitle

\section{Introduction}
Artificial control of {\em direct}
exchange interactions, which occur  at  length scales of one  lattice spacing,
is hardly  possible with current day technologies. In contrast, there is a
number of proposals to control {\em Indirect} Exchange Interactions (IEI) of
spins sitting several nanometers away, that take advantage of the
optical and electrical manipulation of the intermediate fermions afforded  in
semiconducting hosts \cite{AL02,LD98,SI99,PC02,BB02,CD03,FP04}.
Local spins could be provided by  nuclei, by electrons bound to donors
\cite{PC02,BB02}, or $d$ electrons of magnetic impurities \cite{FP04}.
A variety of phenomena, like the reversible modification of the Curie temperature and
coercive fields in (III,Mn)V \cite{OC00} and (II,Mn,N)VI semiconductors, the
induction magnetic order in otherwise paramagnetic (II,Mn)VI semiconductor
quantum dots \cite{MG04} and the entanglement of donor spins in (II,Mn)VI
quantum wells \cite{BB02} have been observed experimentally thanks to
the artificial manipulation of IEI.  Such a control is also required in the
implementation of  quantum computation using localized spins in solids, since two
qbit operations require exchange interactions between distant spin pairs
\cite{Ba95,Di00}.

Laser induced IEI in semiconductor dots with magnetic impurities can be tuned
by changing the laser
frequency, intensity and polarization. For frequencies below the dot gap, an
optical coherence between valence and conduction bands is induced, capable of
mediating exchange interaction between localized spins \cite{PC02,FP04}.
This is commonly known as an Optical RKKY (ORKKY) exchange interaction.
Above threshold (frequencies higher than the dot gap) real carriers (electron-hole pairs)
are responsible for the resulting RKKY-like exchange interaction.
An interesting effect has recently been observed in a system
with two  localized spins interacting  with one itinerant exciton \cite{PQ04}.
By solving exactly this simple case, the authors found that, as the laser energy approaches a
resonance related to excitons bound to impurities, the induced coupling between
spins increases and may switch from ferromagnetic to antiferromagnetic.
On the other hand, the hybridization of the localized and itinerant electrons has
been recently introduced \cite{RL05} concluding that it may produce,
for certain dot geometries, a change of sign in the spin exchange
interaction. Both works point to the same direction: the possibility of
controlling the switching of the exchange interaction to and fro between ferromagnetic (F)
to antiferromagnetic (AF).


Motivated by late experimental results, we have recently  proposed
\cite{CF05} a system where optical exchange interaction is greatly
enhanced:  a semiconductor micropillar cylindric cavity
\cite{SP01,OR04,RS04}, made of CdTe with inclusions of (Cd,Mn)Te
quantum dots \cite{MB00}.  Fine tuning of the cavity modes has been
recently achieved, for instance, by using length tunable
microcavities \cite{FV05} or  photonic crystal membrane nanocavities
\cite{BH05}. Detuning can also be varied by making use of the
temperature dependence of the exciton transition \cite{KM01}. The Mn
spins are exchange-coupled both to   electrons and holes  confined
in the quantum dot\cite{MB00}.  In turn,  quantum dot electron hole
pairs (excitons) are coupled to  the  confined photon field. Pillar
microcavities based on CdMnTe/CdMgTe heterostructures have been
fabricated \cite{OR04} featuring  {\em strong coupling} between 2D
excitons with 0D photons with a Rabi energy  as high as 16 meV,
which can hardly be obtained with cw lasers. The strong coupling
regime between InGaAs quantum dot excitons and 0D  photons in a GaAs
0D cavity  has also recently been achieved, with a Rabi energy of
$\simeq 0.1$ meV \cite{RS04}. This model system, where both photons
and fermions have a zero dimensional density of states,  permits the
{\em exact} diagonalization of the Hamiltonian, considering all
degrees of freedom {\em fully quantum  mechanically}. It turns out
that  confinement of both the light and the intermediate fermions
yields an enhancement of the ORKKY interaction that results to be
strongly anisotropic. For photon frequencies below threshold, and at
sufficiently low temperatures, strong ferromagnetic coupling shows
up without a significant increase in exciton density. In addition we
found \cite{CF05} that the interaction mediated by photon-polaritons
is ten times stronger than the one induced by a classical field for
equal Rabi splitting.

The present work is addressed to investigate the consequences of
having a real cavity with a finite quality factor. Calculations for
the case of a (Cd,Mn)Te  quantum dot and realistic values of the
quality factor ($Q>$200, see Ref. {\onlinecite{RS04}}) give  results
very close to those reported in our previous work \cite{CF05}.
However, at smaller $Q$, a sharp resonance shows up that is
manifested by an abrupt change of the exchange interaction between
impurities from ferromagnetic to antiferromagnetic and back to
ferromagnetic. The resonance becomes sharper as the temperature
decreases and may appear at considerably large positive values of
detuning. The switching in sign of the spin-spin interaction is
somewhat similar to that reported in Refs. [\onlinecite{PQ04,RL05}].

\section{Model Hamiltonian}
\subsection{Hamiltonian}
The dot that confines conduction and valence band electrons  has
intra-band level spacing  larger than all the other inter-band
energy scales  \cite{BD95}, so that we only keep the lowest orbital
level in each band. These levels have a twofold spin degeneracy. The
electric field of the lowest cavity mode lies in the plane
perpendicular to the axis of the cylinder. Thus, there are two
degenerate cavity modes, associated to the two possible polarization
states in that plane.  Their energy  $\hbar\omega$  is supposed to
be closed to the quantum dot band gap $E_{g}$. We choose  circularly
polarized cavity modes and, after canonical quantization, the
corresponding photon creation operator is denoted by
$b^{\dagger}_{\lambda}$, where $\lambda=L,R$. When needed, we
associate $L$ with $-1/2$ and $R$ with $1/2$. The  Hamiltonian here
considered  has three terms,
\begin{equation}
{\cal H}={\cal H}_0+{\cal H}_{g}+{\cal H}_{J}.
\end{equation}
\noindent The first describes  decoupled cavity photons and electrons,
\begin{eqnarray}
{\cal H}_0=\sum_{\lambda} \hbar (\omega + {\rm i}\Delta \omega)
b_{\lambda}^{\dagger} b_{\lambda}
+\sum_{\sigma}\left [ \epsilon_v v_{\sigma}^{\dagger}v_{\sigma}
+\epsilon_c c_{\sigma}^{\dagger}c_{\sigma}\right]
\end{eqnarray}
\noindent $c^{\dagger}_{\sigma}$ and $v^{\dagger}_{\sigma}$ create
 conduction and valence band electrons   (${\sigma}$ denotes the spin) with
energies $\epsilon_c$ and $\epsilon_v$,  and orbital wave functions
$\psi_c(\vec{r})$) and  $\psi_v(\vec{r})$ respectively.
We may take, with no loss of generality, $\epsilon_v$=0 and $\epsilon_c$= $E_g$.
$\hbar\Delta \omega$ is the inverse of the photon lifetime and is a characteristic of the
microcavity. The microcavity quality factor  is commonly defined as
\begin{equation}
Q=\omega/\Delta \omega.
\end{equation}

The light-matter coupling Hamiltonian is:
\begin{eqnarray}
{\cal H}_{g}= g\sum_{\lambda,\sigma}\left(b_{\lambda}^{\dagger}+b_{\lambda}\right)
\left[c_{\sigma}^{\dagger}v_{\sigma}+v^{\dagger}_{\sigma}c_{\sigma}\right]
\delta_{\lambda,\sigma},
\end{eqnarray}
\noindent where we have assumed that there is purely heavy holes.
This assumption
 leads to  the standard spin selective coupling \cite{PC02,FP04,OO84}
that associates photon polarization and fermion spin degrees of freedom.
This kind of coupling breaks spin rotational invariance and privileges the axis of
the cavity,  $\hat{z}$.
\begin{figure}
\includegraphics[width=2.8in,height=2.8in]{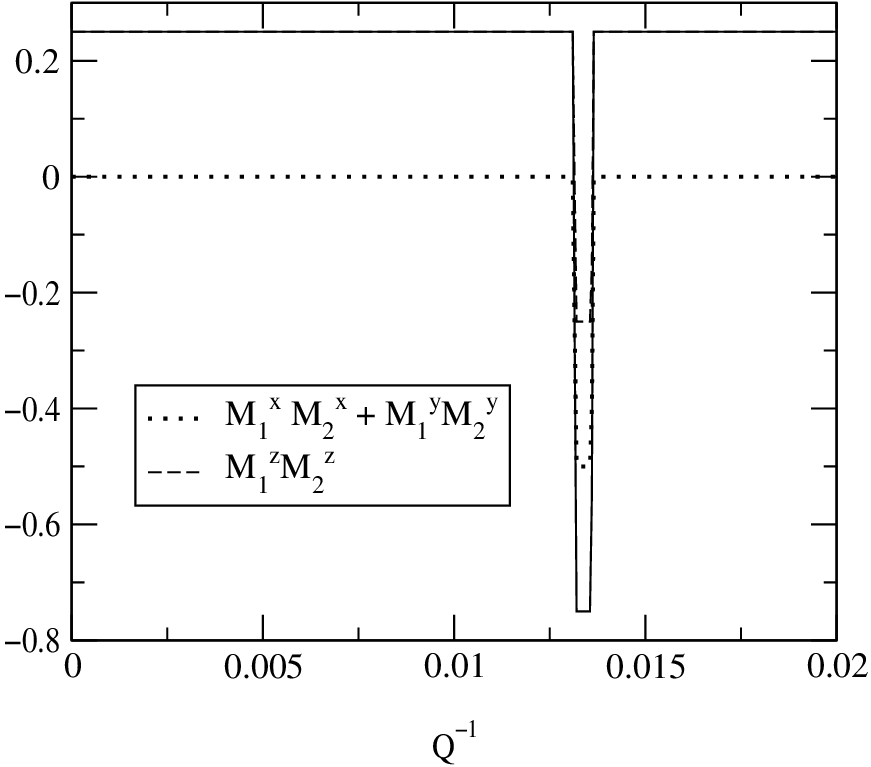}
\caption{Spin-spin correlation functions (continuous line depicts total
correlation ${\vec M}_1 \cdot   {\vec M}_2$) versus the inverse of the
quality factor $Q^{-1}=\Delta \omega/\omega$ for spin 1/2
impurities.
The results correspond to T=0.001 K, $\delta=10$ meV, $g$=5 meV,
and valence and conduction band exchange couplings of
$J_v$ = -0.5 meV and $J_c$= -0.1 meV, respectively.}
\label{spin1.fig}
\end{figure}

The exchange interaction between the fermions and  the spin $M=5/2$ of the Mn
impurities (results for spin 1/2 impurities will also be presented) reads:
\begin{equation}
{\cal H}_{J}= \sum_{I,f}J_f \vec{M}_I\cdot\vec{S}_f\left(\vec{x}_I\right)
\end{equation}
where $\vec{S}_f\left(\vec{r}_I\right)$ stands for local spin density of the
$f=v,c$ electron and $\vec{M}_I$,  is the  Mn spin located at $\vec{r}_I$.
($I=1,2$, two impurities will be hereafter considered).
The electron spin density is,
\begin{equation}
\vec{S}_f\left(\vec{r}_I\right)=\frac{1}{2}|\psi_f(\vec{r}_I)|^2
c^{\dagger}_{\sigma}c_{\sigma'}\vec{\tau}_{\sigma,\sigma'},
\end{equation}
\noindent where $\vec{\tau}$ are the Pauli matrices. The strength of the
interaction  depends
both on the exchange constant of the material $J_f$ and on the localization
degree of the carrier, $|\psi_f(\vec{r}_I)|^2$.

The eigenstates of this Hamiltonian can be classified using the total matter spin
$\vec{\Sigma}_T =\vec{S}_v +\vec{S}_c +\vec{M}_1 +\vec{M}_2$ , and
its $z$-component $\Sigma^z_T= M^z_1+M^z_2+S^z_c+S^z_v$.

We define a spin-spin correlation:
\begin{equation}
\langle\vec{M}_1\cdot\vec{M}_2\rangle= \frac{1}{Z}\sum_{i}
\langle\Phi_i|\vec{M}_1\cdot\vec{M}_2|\Phi_i\rangle {\rm e}^{-E_i/k_BT}
\label{corr}
\end{equation}
\noindent
where $Z$ is the partition function and  the sum runs over the
eigenstates $\Psi_i$ of the Hamiltonian having a real part of the energy $E_i$.

\subsection{Model parameters}
The value of the light-matter interaction  $g$ depends on the amplitude of the
cavity mode in the dot and plays the same role than the Rabi
energy $\Omega$ in the case of a photoexcited semiconductor \cite{PC02,FP04}.
We take $g=5$ meV which is within the range of Rabi splittings reported in the
literature for CdTe nanopillars
(larger values have been reported experimentally \cite{OR04}).
A key quantity of the model is the detuning $\delta = E_g -\hbar\omega$.
As this can be varied experimentally, it will be one of the main variables in
our analysis.

We consider a hard wall quantum dot, with lateral dimensions $L\simeq 10$ nm
and total volume $\simeq 1200$  nm$^3$.  In such a dot, a realistic value  for
the exchange interaction between a conduction (valence) band  electron  and
a Mn spin is $J^{max}_c=-0.1$meV ($J^{max}_v=-0.5$meV ).
The band gap in (Cd,Mn)Te  is the largest energy scale (we take $E_g =$ 2 eV).
Therefore the effect of the terms that do not conserve the number of
excitons plus photons is  negligible and they can safely be removed.
This permits to work in subspaces with ${\cal N}$ excitations.
Here we consider the  coupling between two Mn impurities,
in presence of ${\cal N}=1$ polaritons.
For ${\cal N}=1$ excitation the  ground state manifold is mainly
photonic for $\delta>>0$, mainly excitonic   for $\delta<<0$
and it is a compensated mixture around $\delta=0$ .

Aiming to attain a full understanding of the dependence of the results
on the model parameters, the realistic values for the exchange
interaction and that chosen for the light-matter coupling $g$,
were not always used in the calculations discussed hereafter.
\begin{figure}
\includegraphics[width=2.8in,height=2.8in]{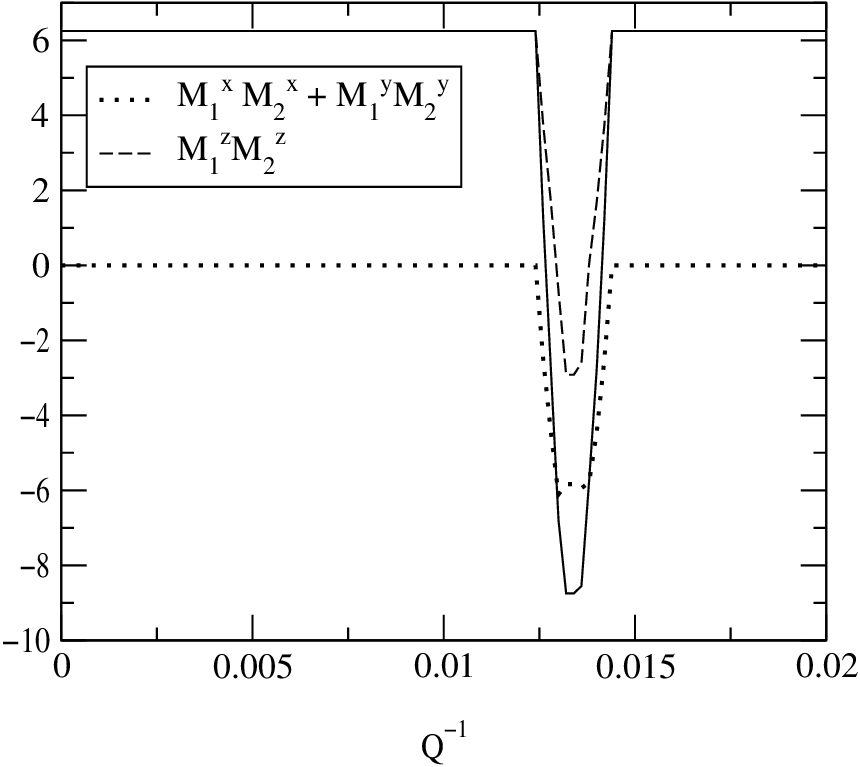}
\caption{Same as Fig. 1 for spin 5/2 impurities.}
\label{spin5.fig}
\end{figure}

\section{Results}
In the following we present results  obtained by varying either
$\Delta\omega$ (and, thus, the quality factor) while keeping
constant the photon frequency (subsection A) or the detuning
$\delta$ (subsection B). The latter will be varied by changing the
photon frequency, the most feasible way in present experimental set
ups. Note that changing detuning over a narrow range around
threshold ($\delta=0$) will also slightly modify the quality factor.
Identical results are obtained if, instead, the dot gap is varied, a
procedure that keeps constant $Q$.

\subsection{Switching  the sign of the exchange interaction by varying the photon lifetime}
\begin{figure}
\includegraphics[width=2.6in,height=2.8in]{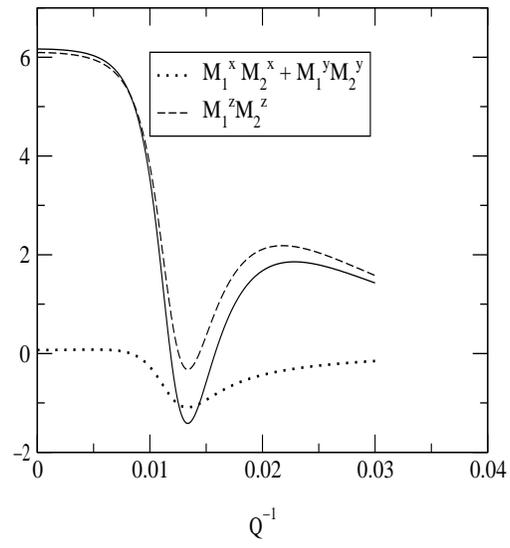}
\caption{Same as Fig. 1 for spin 5/2 impurities and T=0.1 K.}
\label{T0.1K.fig}
\end{figure}

Figs. 1 and 2 depict results for total $\langle M_1 \cdot
M_2\rangle$, out-of plane $\langle M_1^{z}M_2^{z}\rangle$ and
in-plane $\langle M_{1p}M_{2p}\rangle= \langle M_1^{x}M_2^{x}+
M_1^{y}M_2^{y}\rangle$ spin-spin correlations versus the inverse of
the quality factor $Q$ for spin 1/2 and 5/2, respectively, and  a
temperature of $T$ =0.001 K (the rest of the model parameters are
given in the captions).  At large  quality factors the spin-spin
correlation is, in both cases, maximum and fully out-of-plane
\cite{CF05}. In addition, for quality factors larger than 100 no
differences are noted with respect to the case of an ideal cavity
\cite{CF05}. Abrupt changes in the character of the effective
exchange interaction between impurities are noticed for
$Q^{-1}\approx 0.013$ for both spin 1/2 and 5/2 impurities. The
spin-spin correlation abruptly changes from ferromagnetic to
antiferromagnetic and back to ferromagnetic over a narrow range of
$Q$. Ferromagnetic correlations occuring at small $Q$ are fully
in-plane. At the very low temperature as shown in Fig. 1, the
spin-spin correlation reaches always its maximum (ferro or
antiferro) values. The range of quality factor over which this
switching of the sign of the exchange interaction occurs, is
narrower for spin 1/2 impurities. Results for a higher temperature
$T=0.1 K$ for spin 5/2 impurities are shown in Fig. \ref{T0.1K.fig}.
Now, the ideal cavity (infinite quality factor) limit is reached
only at  quality factors larger than 200 \cite{CF05}. In addition,
the abrupt switching observed at the lower temperature is
appreciably smeared. Moreover, although the maximum
antiferromagnetic correlation occurs at roughly the same quality
factor, it is sharply weakened. Altogether, the spin correlation
decreases steadily with the inverse of the quality factor. This
decrease is in fact what one should expect at any temperature.
However at very low temperatures the spin correlation remains
constant up to rather low values of $Q$ for the reasons discussed
hereafter.

In order to understand the origin of the behavior of spin-spin
correlations, and in particular of the switching mentioned above, we
proceed to analyze how the ground and excited states evolve, in the
simple case of spin 1/2 impurities, as the quality factor is
decreased. Let us denote $|\lambda;\uparrow
\downarrow,0;M_1^z,M_2^z>$ and $|0;S^z_v,S^z_c;M_1^z,M_2^z>$
photonic and excitonic configurations with $z$-components of the two
spin impurities $M_1^z$ and $M_2^z$ and either one photon of
polarization $\lambda$ (we denote right polarization by $\uparrow$
and left polarization by $\downarrow$) and two electrons in the
valence band or no photon and two electrons one in the valence band
and other in the conduction band of $z$-component of the spin
$S^z_v$ and $S^z_c$, respectively. The spin-selective coupling
considered here requires that $S^z_c=\lambda$. At infinite quality
factor  and in the absence of exchange coupling, the ground state is
eight-fold degenerate. Exchange coupling lifts this degeneracy as
follows. Two of these states, both doubly degenerate, give
ferromagnetic correlation among impurities,
\begin{subequations}
\begin{equation}
|\psi_{-1}>=a_{-1} |\uparrow;\uparrow\downarrow,0;\downarrow,\downarrow> +
b_{-1} |0;\downarrow,\uparrow;\downarrow,\downarrow>.
\end{equation}
\begin{equation}
|\psi_{1}>=a_{1} |\uparrow;\uparrow\downarrow,0;\uparrow,\uparrow> +
b_{1} |0;\downarrow,\uparrow;\uparrow,\uparrow>,
\end{equation}
\end{subequations}
\noindent Both states may involve other basis functions albeit with
a very small weight. The quantum numbers of these states are
$\Sigma_{T}^z$=-1 and $\Sigma_T = 1$ for $|\psi_{-1}>$ and
$\Sigma_{T}^z$=1 and  $\Sigma_T= 1$ for $|\psi_1>$. They are
degenerate with the states corresponding  to a linear combination of
left polarization $\lambda = \downarrow$ and spin down electron in
the conduction band.

The  third state is also doubly degenerated and give antiferromagnetic correlations,
\begin{eqnarray}
|\psi_0>&=&a_0 \left[|\uparrow;\uparrow\downarrow,0;\uparrow,\downarrow> -
|\uparrow;\uparrow\downarrow,0;\downarrow,\uparrow>\right]
\pm\left(|\downarrow;\uparrow\downarrow,0;\downarrow,\uparrow>-
|\downarrow;\uparrow\downarrow,0;\uparrow,\downarrow>\right) \nonumber \\
&&+ b_0 \left[|0;\downarrow,\uparrow;\uparrow,\downarrow>-
|0;\downarrow,\uparrow;\downarrow,\uparrow>\right]
\pm\left(|0;\uparrow,\downarrow;\uparrow,\downarrow>-
|0;\uparrow,\downarrow;\downarrow,\uparrow>\right)
\end{eqnarray}
\noindent This state has total spin $ \Sigma_{T}$ = 0. Therefore, left and right photon
polarizations occur with equal weight. Symmetric and anti-symmetric combinations of
basis functions with opposite photon polarizations are degenerate.
The remaining two states are not degenerate and  have energies very close to that of
$|\psi_0>$  \cite{CF05}.
In these states the total spin is $ \Sigma_T \ne 0$, but $ \Sigma_T^z $ = 0.
Left and right polarizations are equally mixed,  but symmetric and anti-symmetric
combinations of the photon polarizations are no longer degenerate,  because
the state with photon polarization antiparallel to the total spin $\vec{\Sigma}_T$
has lower energy than that having photon  polarization parallel to the total spin (see below).

The way the eightfold degeneracy is lifted when  the exchange
Hamiltonian ${\cal H}_J$ is switched on,
is better understood by rewriting ${\cal H}_{J}$ as,
\begin{equation}
{\cal H}_{J}= J_v(\vec{M}_1+\vec{M}_2) \cdot ({\vec{S}_v}+{\vec{S}_c})+
(J_c-J_v)(\vec{M}_1+\vec{M}_2) \cdot {\vec{S}_c}.
\end{equation}
\noindent It is clear that this Hamiltonian does not change the
energy of states that only involve basis functions in which the two
impurities have opposite spins (as is the case of the state of Eq.
(9)). Instead the energies of the first two states are modified in
first order. Degeneracy is lifted  as diagonal elements from the
second term  in ${\cal H}_{J}$  give non zero contributions. If
$J_c-J_v > 0$ (as is the case of the realistic model parameters
given above), the two states in which the two impurities have
parallel spins antiferromagnetically (ferromagnetically) coupled
with  the photon and the conduction band electron, reduce (increase)
their energy (see Fig. 2a right). Upon switching on the exchange
interaction the three states of Eqs. (8) and (9) keep their
spin-related twofold degeneracy. This is most clearly seen by
writing a  secular equation which is valid for the three
wavefunctions,
\begin{equation}
\left[E+\delta-{\rm i}\Delta\omega\right]\left[E-k(J_c-J_v)\right]-g^2=0
\end{equation}
\noindent where the energy $E$ is referred to the dot band gap
$E_{g}$, and $k$ runs over the subindexes of the three wave
functions, namely, $k=-1,0,1$. When exchange coupling is zero the
three wavefunctions are degenerate, while degeneracy is lifted when
it is switched on.  Note that this equation approximately gives the
energy of the eigenstates in Eqs. (8) and (9) and those of three
additional eigenstates that lie at higher energies \cite{CF05}.

Fig. 4 shows the real part of the energy of these states versus the inverse of the
quality factor $Q^{-1}$ for two sets of values of the light-matter and
exchange couplings. Two state crossings are clearly visible (see insets in Fig. 4). At low
$Q^{-1}$, the ground state is $|\psi_{-1}>$, whereas for large $Q^{-1}$, state
$|\psi_1>$ is the one with the lowest energy. Around the value at which
maximum antiferromagnetic correlation shows up, the ground state
is $|\psi_0>$.  The question now is why these crossings occur.
Adding an imaginary part to the photon frequency increases the
energy of all states involving photonic configurations. However, the
actual increase {\it depends} on how close to the photon energy these states lie.
This offers a qualitative explanation of  the crossings of Fig. 4.

As regards the quantitative results shown in Fig. 4  the following
features are worth commenting.  Increasing the exchange coupling
enlarges the range of quality factors over which level crossings
occur, although the first crossing shows up at roughly the same $Q$.
In addition, energy differences increase suggesting a weaker
dependence on temperature. Increasing light-matter coupling keeping
constant the exchange coupling, does also enlarge energy differences
and decreases the $Q$ at which crossings occur. The latter is due to
the fact that, as decreasing the quality factor decreases the photon
density of states, a larger light-matter coupling is required to
produce the same effect. This is an interesting effect which
suggests that in order to produce the switching in sign of the
exchange interaction in cavities with a sufficiently long photon
lifetime, light-matter coupling should be decreased. On the other
hand, effects of couplings on energy differences can be trivially
understood in terms of the secular equation given above.

An aditional feature of the results shown in Fig. 4 is important to
comment. For large quality factors the state in which the spin of
the impurities is antiparallel to the photon polarization has the
lowest energy. However, once crossings have occurred, the state
having the spin impurities parallel to the photon polarization
becomes the ground state. This interesting effect opens the
possibility of switching the photon polarization as follows. As
remarked above the three states are twofold degenerate. This
degeneracy can be removed by applying a weak magnetic field. Then,
as the magnetic field forces the spin of the impurities to lie along
the field, the above crossing will imply an inversion of photon
polarization. Note that, if the magnetic field is sufficiently weak,
level crossings will not be eliminated. The field, however, cannot
be too low, otherwise a very low temperature will impede this effect
to show up. The effect can be most easily produced by changing
detuning (see below).

\subsection{Varying the photon frequency}

In the previous section we have shown how changing the cavity quality factor
induces a switching in the sign of the exchange interaction. However, the quality
factor is not a parameter that can be easily varied. In this section we show that
sign switching can also be triggered by varying detuning  \cite{FV05,BH05}.
Detuning will be varied by changing the photon frequency.
Figs. 5 and 6 show  spin-spin correlations and  exciton and photon occupation numbers,
versus detuning $\delta$, for two values of the quality factor
$Q\sim100$ and $Q\sim200$, respectively.
Note that as $\delta$ is  usually varied over a narrow range around $\delta =0$,
the quality factor $Q$  changes only slightly ($\omega$ remains always close to $E_g$) \cite{note}.
In order to get the switching of the exchange interaction at a
sufficiently large quality factor, we take $g = 2.5$ mev. As such low light-matter
couplings excessively reduce the energy differences between the states
that matter (see above), we
also choose the largest values of the exchange couplings used to obtain the results
presented in Fig. 3, namely, $J_c=-0.2$meV and $J_v=-1.$meV.

As detuning is varied, we go continuously from a pure photonic state ($\delta > 0$) to
a pure excitonic state ($\delta < 0$). For a cavity with an infinite quality
factor, the exciton and photon occupation numbers  vary smoothly with
$\delta$ \cite{CF05}. Besides, the crossover from a photonic to an excitonic
state occurs at  $\delta=0$ \cite{CF05}.
Finite quality factors modify this behaviour because  the light-matter
interaction is weakened, shifting also that crossover to positive detuning (see
Figs. 5b and 6b).  For $Q\sim 100$ (Fig. 5), the crossover occurs over a
very narrow window of positive detuning values. For a larger
$Q \sim 200$ (see Fig. 6) the effective light-matter coupling  is raised up and
therefore the transition from the photonic to the excitonic state is less
abrupt and occurs at a lower detuning.
Note that, on the pure photonic side, correlations (either ferromagnetic
or antiferromagnetic) are large, with a neglegible exciton population,
a fact already observed in our previous work \cite{CF05}. The latter
effect is here enhanced due to the efective reduction of the electron-photon
interaction produced by a finite quality factor.


As discussed in the preceding subsections, at high quality factors the AF-F
transition is shifted to lower $\delta$. Thus, the
$\delta$ value at which  the ground state switches from photon--like
to exciton--like may be approached.
When that switching occurs, the AF peak dissapears abruptly (see Fig. 6),
revealing the fundamental role played by
the photon in inducing correlation between impurity spins.
Another important point is that, at  large quality factors, the effect of temperature
on the AF peak is weakened.
This can be understood by remembering that the energy differences between
the three states of Eqs. (8) and (9)
are proportional to the effective exchange interaction, which is raised up
as detuning decreases \cite{CF05}. As  shown in  Fig. 6,   increasing the temperature
up to $T = 0.1$ K scarcely affects the AF-F transition; moreover the maximum AF
spin-spin correlation is still reached.
The width of the frequency window in which the AF correlations rise up can be
enlarged by changing the
couplings $J_v$ and $J_c$. For $Q=200$, with $J_c= -0.085$ meV and $J_v=0.1$ meV
a window as large as 5 mev is obtained.  Finally note that,
the effect highlighted above, related to a possible change of photon
polarization induced by a weak magnetic field, may also appears when
detuning is varied. This is definitively proved by the numerical
results shown in Fig. 7. For large positive detuning the correlation
between the spin of the impurities and the photon polarization is
negative. This correlation function changes smoothly over the
detuning range where the F-AF-F transition occurs, going through
zero as expected for two fully AF correlated impurities. Beyond the
transition, the spin-photon polarization correlation becomes
positive.



\section{Concluding Remarks}
Summarizing, we have studied the indirect exchange interaction
between two spins in a cavity-dot system with one polariton,
assuming that the cavity has a finite quality factor. For the best
cavities nowadays available (quality factors better than 200), we
have shown that our previous results on ideal cavities (infinite
cavity factor) \cite{CF05} remain valid. However,  if the quality
factor is decreased a switching of the exchange interaction to and
fro between ferromagnetic and antiferomaggnetic may be the case. We
were able to demonstrate that this effect occurs for realistic
values of the model parameters and does not require excessively low
temperatures. The effect can be experimentally proved by varying
detuning, a parameter that can be reliably varied in several
experimental configurations available nowadays. In addition we have
shown that this switching is related to level crossings in such a
way that the ground state changes from one in which the impurity
spins are AF correlated to the conduction band electron, to one in
which this correlation is just the opposite. We have suggested that
this may induce inversion of the photon polarization with the help
of a small magnetic field. Although more work is of course required
to fully understand this syatem, we believe that the results
presented here clearly illustrate the possibility of switching the
sign of the exchange interaction,  an effect that has recently
attracted a considerable interest \cite{PQ04,RL05}.


\acknowledgments
We acknowledge several interesting discussions with F. Fern\'andez-Rossier
during the first stages of this work.
Financial support by the spanish MCYT (grants MAT2005-07369-C03-01 and
NAN2004-09183-C10-08), the Universidad de Alicante, the Generalitat Valenciana
(grant GRUPOS03/092 and grant GV05/152), the Universidad de Buenos
Aires (grant UBACYT x115),  the argentinian CONICET, and the the brazilian agencies
FAPERJ, CAPES and CNPq, is gratefully acknowledged. GC is
thankful to the spanish "Ministerio de Educaci\'on y Ciencia" for a Ram\'on
y Cajal grant.

\begin{figure}
\includegraphics[width=6in,height=6in]{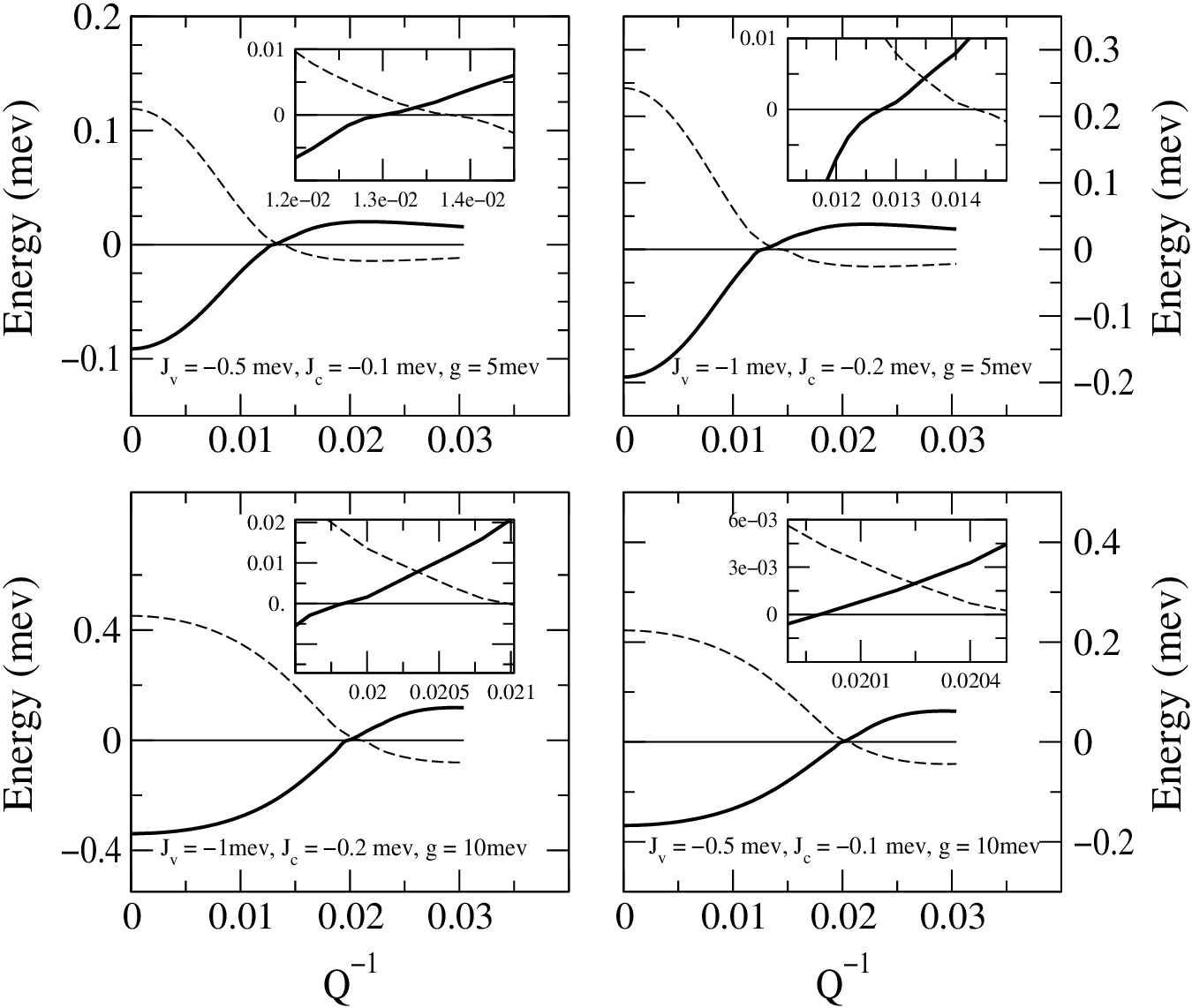}
\caption{Real part of the energies of the states $|\psi_{-1}>$ (thick continuous line) $|\psi_1>$
(broken line) and $|\psi_0>$ (thin continuous line) of
Eqs. (8a), (8b) and (9) respectively,
versus the inverse of the quality factor $Q^{-1}=\Delta \omega/\omega$.
The results correspond to spin 1/2 impurities,and $\delta=10$ meV and
values of the light matter  coupling $g$ and valence and conduction band exchange
couplings ($J_v$  and $J_c$) given in the Figures.
All energies referred to the energy of state $|\psi_0>$.}
\end{figure}
\newpage
\begin{figure}
\includegraphics[width=3.5in,height=4.5in]{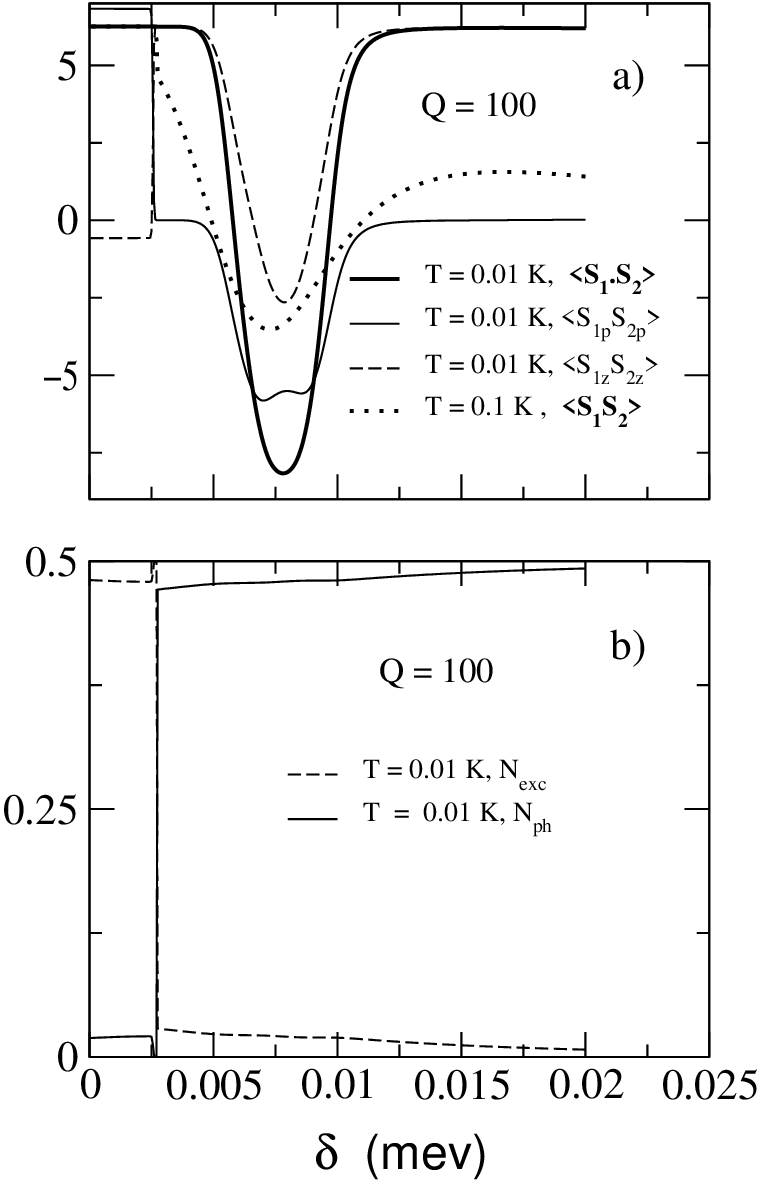}
\caption{a) Spin correlation functions versus detuning $\delta$ for spin
5/2 impurities. b) Photon and exciton occupation numbers.
The results correspond to $Q$=100,  $g$=2.5 meV,
and valence and conduction band exchange couplings of
$J_v$ = -1 meV and $J_c$= -0.2 meV, respectively.}
\end{figure}

\newpage
\begin{figure}
\includegraphics[width=3.5in,height=4.2in]{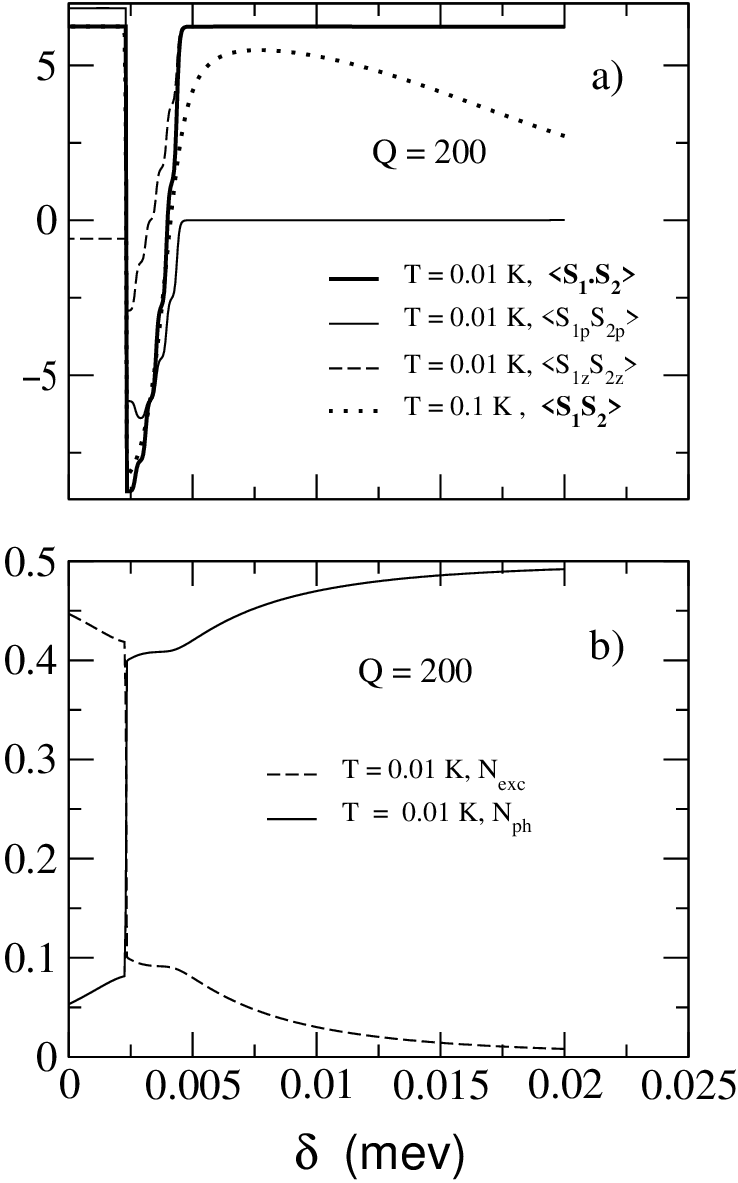}
\caption{Same as Fig. 4 for $Q$=200}
\end{figure}

\newpage
\begin{figure}
\includegraphics[width=3in,height=2.3in]{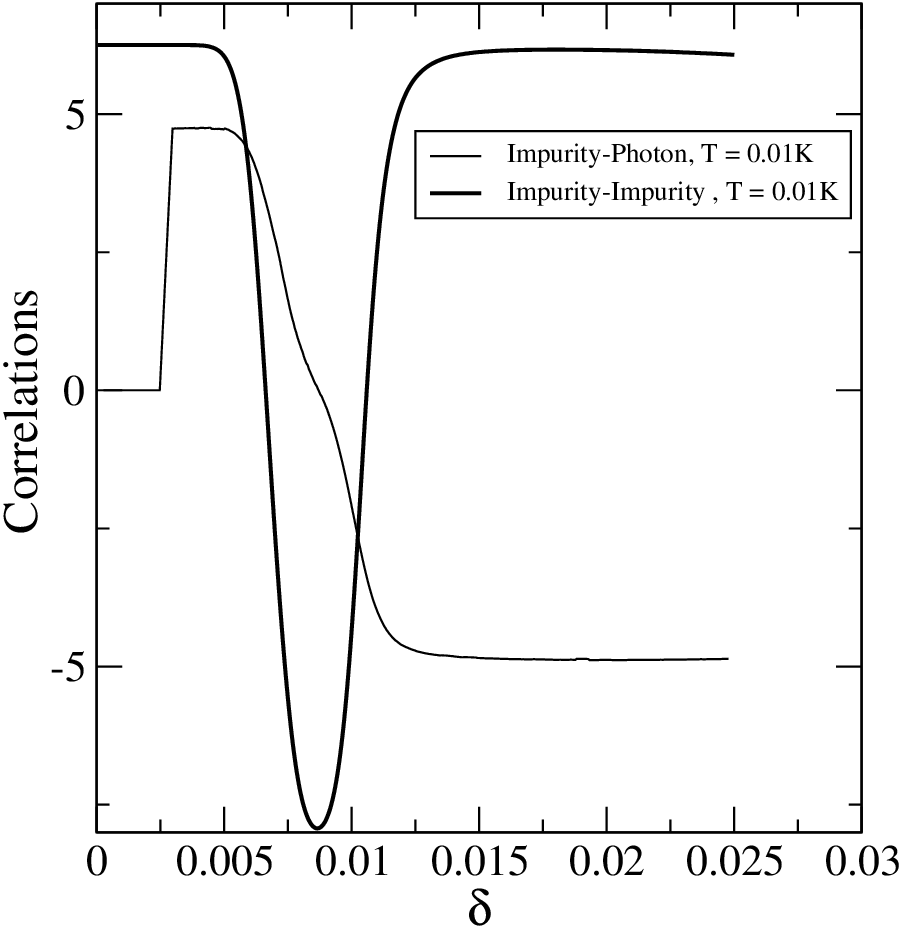}
\caption{Spin-spin and photon polarization-total impurity spin  correlation functions versus
$\delta$ for spin 5/2 impurities for the parameters of Fig. 5.}
\end{figure}

\begin{thebibliography}{99}
\bibitem{note}
We are assuming that the photon lifetime does not depend on the photon
frequency, and, thus, the quality factor is linearly related
to the latter. Note, however, that this may  not always be the case.
\end{thebibliography}
\end{document}